\documentclass{article}

\usepackage[final]{neurips_2023}

\usepackage[utf8]{inputenc} 
\usepackage[T1]{fontenc}    
\usepackage{hyperref}       
\usepackage{url}            
\usepackage{booktabs}       
\usepackage{amsfonts}       
\usepackage{amsmath}
\usepackage{nicefrac}       
\usepackage{microtype}      
\usepackage{xcolor}         
\usepackage{afterpage}
\usepackage{subcaption}
\usepackage{float}
\usepackage{graphicx}
\usepackage{caption}
\newcommand{\mean}[1]{\left\langle #1 \right\rangle}
\newcommand{\bs}[1]{\boldsymbol{#1}}
\newcommand{\deriv}[2]{\frac{\partial#1}{\partial #2}}

\title{Physics-Informed Tensor Basis Neural Network for Turbulence Closure Modeling}

\author{%
    Leon Riccius, Atul Agrawal \& Phaedon-Stelios Koutsourelakis \\
  Professorship of Data-driven Materials Modeling\\
  School of Engineering and Design\\
  Technical University of Munich\\
  Boltzmannstr. 15, 85748 Garching, Germany \\
  \texttt{\{leon.riccius, atul.agrawal, p.s.koutsourelakis\}@tum.de}\\
}

\begin{document}

\setlength{\abovedisplayskip}{-12pt}
\setlength{\belowdisplayskip}{-3pt}
\setlength{\textfloatsep}{0.075cm}

\maketitle

\begin{abstract}
Despite the increasing availability of high-performance computational resources, Reynolds-Averaged Navier-Stokes (RANS) simulations remain the workhorse for the analysis of turbulent flows in real-world applications.
Linear eddy viscosity models (LEVM), the most commonly employed model type, cannot accurately predict complex states of turbulence.
This work combines a deep-neural-network-based,  nonlinear eddy viscosity model with turbulence realizability constraints as an inductive bias
in order to yield improved predictions of the anisotropy tensor. 
Using visualizations based on the barycentric map, we show that the proposed machine learning method's anisotropy tensor predictions offer a significant improvement over all LEVMs in traditionally challenging cases with surface curvature and flow separation. 
However, this improved anisotropy tensor does not, in general, yield improved mean-velocity and pressure field predictions in comparison with the best-performing LEVM.
\end{abstract}

\section{Introduction}

The incompressible Navier-Stokes equations are vital for describing fluid motion at low Reynolds numbers, impacting fields like aircraft design and ocean current modeling.
Turbulent flows are prohibitively expensive to resolve fully, and engineers often resort to reduced models such as  RANS for efficiency.
These models employ closures such as the Launder-Sharma $k-\epsilon$ \cite{Launder1974} or Wilcox's $k-\omega$ \cite{Wilcox2008}, which rely on linear assumptions, limiting their accuracy in complex flow scenarios.
Nonlinear models have been explored but face challenges.
This contribution builds upon the resurgence of turbulence modeling research, instigated by data-driven approaches \cite{Duraisamy2019}, in response to the stagnation seen in the 2000s after earlier advancements.

This work combines additional flow features derived by Wang and colleagues \cite{Wang2016, Wang2017} with the neural network architecture proposed by Ling et al. \cite{Ling2016} to give point-based estimates of the anisotropy tensor appearing in the RANS closure.
The training objectives are supplemented by a loss term penalizing predictions that violate the physical realizability constraints of turbulent states.
The trained network proposed was tested on unseen flow scenarios and used as a source term in the RANS equations to produce estimates of the mean-flow quantities.

\section{Physics-Informed Tensor Basis Neural Network}

\subsection{Reynolds stresses and realizability constraints}

The Reynolds stress tensor $\tau_{ij} = \mean{u_i u_j}$, where $u_i = U_i - \mean{U_i}$ arises by time-averaging (mean indicated by $\langle \cdot \rangle$) of the Navier-Stokes equations and depends on the fluctuations $u_i$ of the velocity field $U_i$.  It can be decomposed into an isotropic $\delta_{ij}$ and anisotropic $a_{ij}$ part which is given by $a_{ij} = \tau_{ij} - \nicefrac{2}{3} k \delta_{ij}$.The turbulent kinetic energy $k$ is the trace of the Reynolds stresses.

The anisotropic $a_{ij}$ has zero trace, and its normalized version $b_{ij}$ is referred to as anisotropy tensor, i.e.:

\begin{align}
b_{ij} = \frac{a_{ij}}{2k} = \frac{\tau_{ij}}{\mean{u_k u_k}} - \frac{1}{3} \delta_{ij},
\end{align}

The Reynolds stress tensor is a symmetric, positive semi-definite second-order tensor with a non-negative determinant and trace.
Following \cite{Schumann1977}, the physical constraints, known also as  realizability constraints, on the anisotropy tensor are:

\begin{align}
\label{eq:constraints_b}
    - \frac{1}{3} \leq b_{\alpha \alpha} \leq \frac{2}{3} \quad \forall \alpha \in \{1, 2, 3\},
    \quad \quad -\frac{1}{2} \leq b_{\alpha \beta} \leq \frac{1}{2} \quad \forall \alpha \neq \beta.
\end{align}

\begin{figure}
    \centering
    \begin{subfigure}[b]{0.23\textwidth}
        \includegraphics[width=\textwidth]{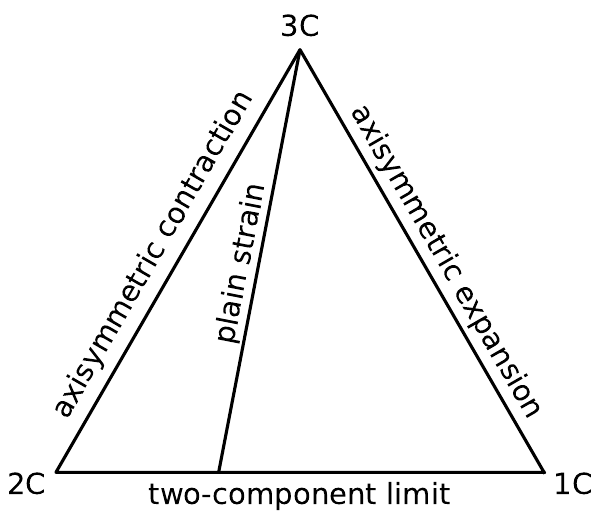}
        \caption{Characteristic regions}
        \label{subfig:barycentric_traingle_characteristic_regions}
    \end{subfigure}
    \begin{subfigure}[b]{0.23\textwidth}
        \includegraphics[width=\textwidth]{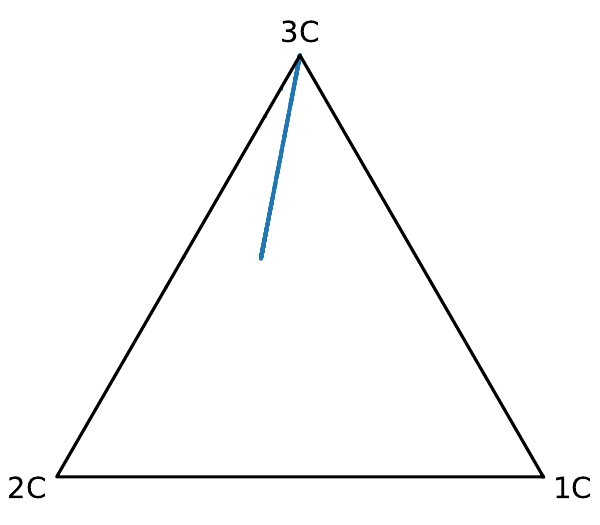}
        \caption{RANS $k-\omega$}
    \end{subfigure}
    \begin{subfigure}[b]{0.23\textwidth}
        \includegraphics[width=\textwidth]{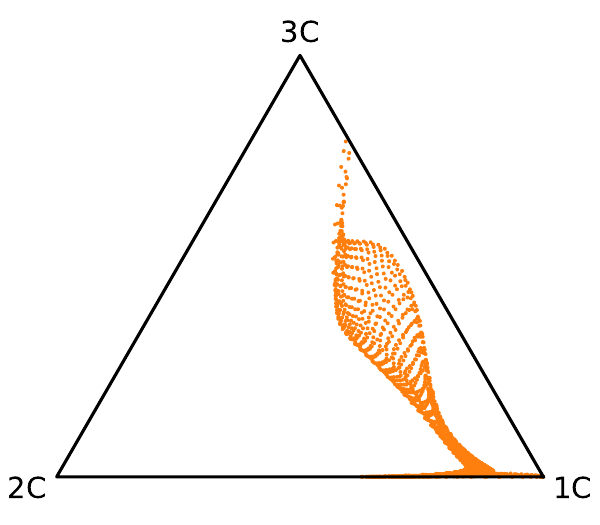}
        \caption{DNS}
    \end{subfigure}
    \begin{subfigure}[b]{0.23\textwidth}
        \includegraphics[width=\textwidth]{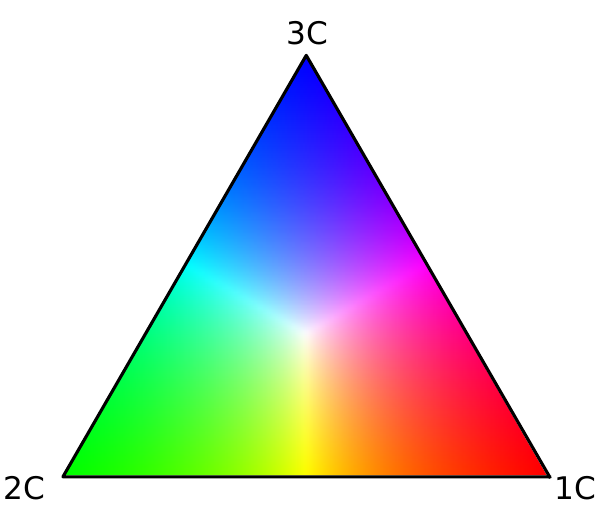}
        \caption{RGB mapping}
    \end{subfigure}
    \caption{Barycentric map representing nature of turbulence of RANS $k - \omega$ \textbf{(a)} and DNS (b)}
    \label{fig:rgb_map_scaled}
\end{figure}

The barycentric map introduced in \cite{Banerjee2007} uses an eigenvalue decomposition of $\bs{b}$ to define three fundamental states of turbulence.
All other states of turbulence can be expressed as a linear combination of these three limiting states.
The limiting states form a triangle of all admissible states;
its vertices are defined by the realizability constraints \eqref{eq:constraints_b}.
As demonstrated in Figure \ref{fig:rgb_map_scaled}, each point in the barycentric triangle corresponds to a unique color. The mapping is given in the supplementary material \ref{subsec:colormap}.


\subsection{Closure Model}

While LEVMs assume $\bs{b}$ to be a linear function of the mean velocity gradient, a more general class of turbulence models can be formulated when dropping this assumption.
The class of algebraic stress models is formed by nonlinear eddy viscosity models (NLEVM), which determine the Reynolds stresses from the local turbulent kinetic energy $k$, the eddy viscosity $\epsilon$, and the mean velocity gradient.
Pope \cite{Pope1975} has shown that every second-order tensor that can be formed from the normalized mean rate of strain $\hat{\bs{S}} = \nicefrac{\epsilon}{2k} (\nabla \mean{\bs{U}} + \nabla \mean{\bs{U}}^T)$ and the normalized rate of rotation $\hat{\bs{\Omega}} = \nicefrac{\epsilon}{2k} (\nabla \mean{\bs{U}} - \nabla \mean{\bs{U}}^T)$ and fulfills these requirements is a linear combination of ten basis tensors $\mathcal{T}_{ij}^{(n)}$.
The most general form of a  NLEVM is given by

\begin{align}
\label{eq:nonlinear_eddy_viscosity_model}
    b_{ij} = \sum_{n=1}^{10} G^{(n)}(\lambda_1, ..., \lambda_5) \, {\mathcal{T}}_{ij}^{(n)}(\hat S_{ij}, \hat \Omega_{ij}),
\end{align}

where $G^{(n)}$ are the coefficients of the basis tensors and $\lambda_k$ are the tensor invariants dependent on $\hat{\bs{S}}$ and $\hat{\bs{\Omega}}$.
Ling et al. \cite{Ling2016} introduced the Tensor Basis Neural Network (TBNN), which makes use of modern machine learning methods to learn these functions $G^{(n)}$ from high-fidelity fluid simulation data.
Even though improved results compared to the $k - \epsilon$ model were reported, extracting enough information from these five invariants has proven difficult.
This is especially true for flow cases with at least one direction of homogeneity, where invariants $\lambda_3$ and $\lambda_4$ vanish for the entire flow domain.
Proof of this statement is given in the supplementary material \ref{sec:invar_proof}.
It is, however, possible to include more features from local flow quantities and derive more invariants while still employing the integrity basis formed by $\mathcal{T}^{(n)}$.
This work, in part, follows the research of Wang et al. \cite{Wang2018}, who derived an extended feature set also considering the gradients of the turbulent kinetic energy and the pressure. The model was implemented in PyTorch \cite{Paszke2019} and can be accessed via \href{https://github.com/pkmtum/PI_TBNN}{github.com/pkmtum/PI\_TBNN}.

\subsubsection{Enforcing Realizability Constraints in Training}

Ling et al. \cite{Ling2016} enforced the realizability constraints in post-processing by simply projecting the points onto the closest boundary of the barycentric triangle.
We propose incorporating these constraints in the TBNN's training, which we then colloquially refer to as the Physics-Informed Tensor Basis Neural Network (PI-TBNN).
The inequalities given in Eq. \eqref{eq:constraints_b} can be transformed into contributions to the loss function via the penalty method.
The additional term reflects inductive bias about the problem structure and essentially acts as a regularizer. In plain words, if training samples are outside the domain of realizable turbulence states, the penalty term will force them back in.
The constraints are

\begin{equation}
  \begin{split}
    c_1(\bs{b}) &= \min_{\alpha} (b_{\alpha \alpha}) - \nicefrac{1}{3} < 0 \quad \forall \alpha \in \{1,2,3\}, \\
    c_2(\bs{b}) &= ({3 |\phi_2| - \phi_2})/2 - \phi_1 < 0,\\
    c_3(\bs{b}) &= \nicefrac{1}{3} - \phi_2 < 0,
  \end{split}
  \hspace{0.4cm}
  \begin{split}
    c_4(\bs{b}) &= 2|b_{12}| - \left( b_{11} + b_{22} + \nicefrac{2}{3} \right) < 0, \\
    c_5(\bs{b}) &= 2|b_{13}| - \left( b_{11} + b_{33} + \nicefrac{2}{3} \right) < 0, \\
    c_6(\bs{b}) &= 2|b_{23}| - \left( b_{22} + b_{33} + \nicefrac{2}{3} \right) < 0, \\
  \end{split}
\end{equation}
\vspace{0.05cm}

\noindent where $\phi_i$ denotes the eigenvalues of $\bs{b}$ in order to distinguish them from the invariants.
$\phi_1$ and $\phi_2$ are the largest and second-largest eigenvalues.
The penalty term is then given by

\begin{align}
    \mathcal{L}(\hat{\bs{b}}(\bs{\lambda}, \bs{\theta})) = \beta \sum_{k=1}^6 \max(0,c_k(\hat{\bs{b}}(\bs{\lambda}, \bs{\theta}))),
\end{align}

\noindent where $\bs{\lambda}$ is the collection of invariants, $\bs{\theta}$ are the NN parameters, and $\bs{\hat b}(\bs{\lambda},\bs{\theta})$ is the predicted anisotropy tensor. The penalty coefficient $\beta$ determines its impact on the loss function.
The complete loss function, considering the MSE loss, the regularization, and penalty terms, is given by

\begin{align}
    E(\bs{\lambda}_i, \bs{\theta}) &= \frac{1}{D} \sum_{i=1}^{D} \| \bs{\hat b}(\bs{\lambda}_{i}, \bs{\theta}) - \bs{b}_i \|
    + \frac{\alpha}{2} \| \bs{\theta} \|^2_2
    + \frac{\beta}{D} \sum_{i=1}^{D} \sum_{k=1}^6 \max(0,c_k(\bs{\hat b}(\bs{\lambda}_{i}, \bs{\theta}))),
\end{align}

\noindent where $\bs{b}_i$ are the high-fidelity responses. The number of data points is denoted $D$. The coefficient $\alpha$ controls the degree of $L_2$ regularization.

\section{Numerical Results}

A total of four flow geometries were used as benchmarks in this work. The flow over periodic hills (PH) \cite{Mellen2000},  the converging-diverging channel flow (CDC) \cite{Laval2011}, and the curved backward-facing step (CBFS) \cite{Bentaleb2012}. These exhibit adverse pressure gradients over curved surfaces, leading to flow separation and subsequent reattachment.
The data set was also extended to include the square duct (SD) flow case \cite{Pinelli2010}.
This scenario clearly illustrates the limitations of LEVMs and is well-suited for investigating the forward propagation of the predicted anisotropy tensor.
All flow cases were replicated in OpenFOAM \cite{Weller1998} to obtain the baseline RANS data. The flow case setup is analogous to \cite{Riccius2021}.
A detailed description of the data sets is given in the supplementary material \ref{subsec:data_set}.

\subsection{Anisotropy Tensor Prediction}

\begin{figure}
\centering
    \begin{subfigure}[b]{0.028992\textwidth}
        \includegraphics[width=\textwidth]{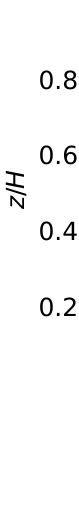}
    \end{subfigure}
    \begin{subfigure}[b]{0.1472\textwidth}
        \includegraphics[width=\textwidth]{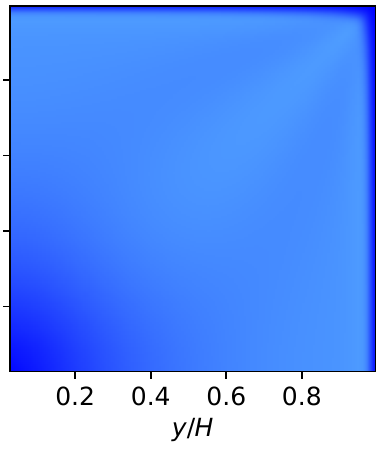}
        \caption{\footnotesize LEVM $k-\omega$}
        \vspace{-0.15cm}
    \end{subfigure}
    \begin{subfigure}[b]{0.1472\textwidth}
        \includegraphics[width=\textwidth]{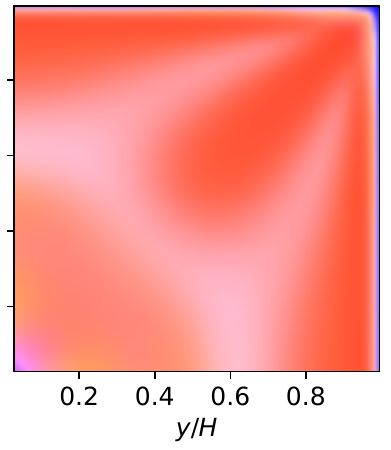}
        \caption{\footnotesize PI-TBNN1}
        \vspace{-0.15cm}
    \end{subfigure}
    \begin{subfigure}[b]{0.1472\textwidth}
        \includegraphics[width=\textwidth]{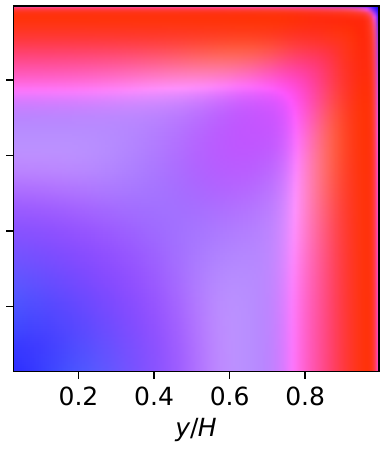}
        \caption{PI-TBNN2}
        \vspace{-0.15cm}
    \end{subfigure}
    \begin{subfigure}[b]{0.1472\textwidth}
        \includegraphics[width=\textwidth]{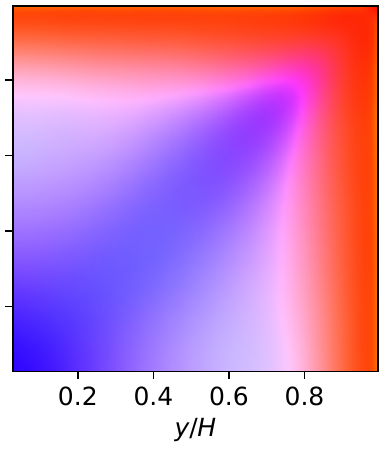}
        \caption{DNS}
        \vspace{-0.15cm}
    \end{subfigure}
    \caption[Visualization of stress types of the $k-\omega$ model, the PI-TBNN with different feature sets, and DNS data for square duct.]{Stress types of $k-\omega$ \textbf{(a)}, PI-TBNN with FS1 \textbf{(b)}, FS[1-3] \textbf{(c)}, and DNS \textbf{(d)} for SD.}
    \label{fig:prediction_b_sd_rgb_map}
\end{figure}

\begin{figure}
\centering
    \begin{subfigure}[b]{0.32\textwidth}
        \includegraphics[width=\textwidth]{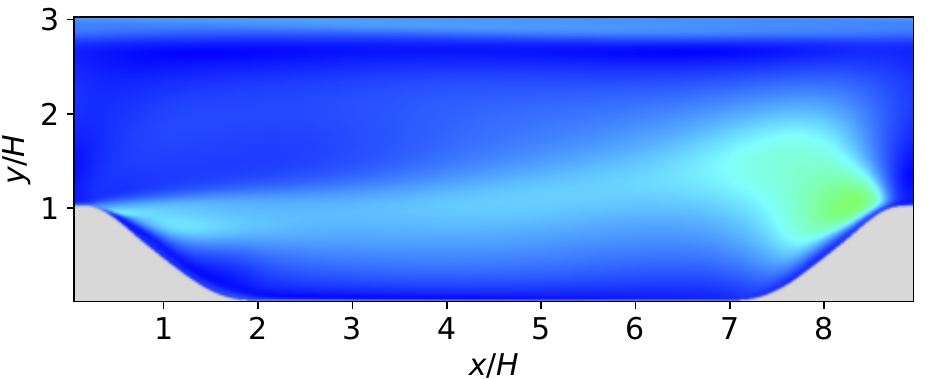}
        \caption{LEVM $k-\omega$}
    \end{subfigure}
    \begin{subfigure}[b]{0.32\textwidth}
        \includegraphics[width=\textwidth]{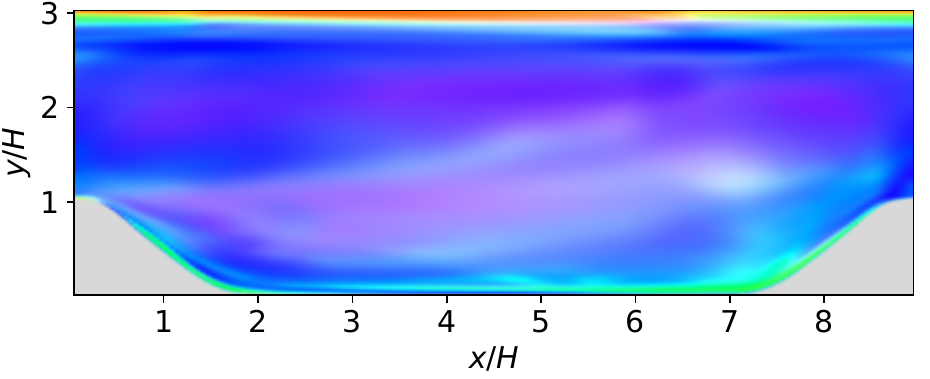}
        \caption{PI-TBNN}
    \end{subfigure}
    \begin{subfigure}[b]{0.32\textwidth}
        \includegraphics[width=\textwidth]{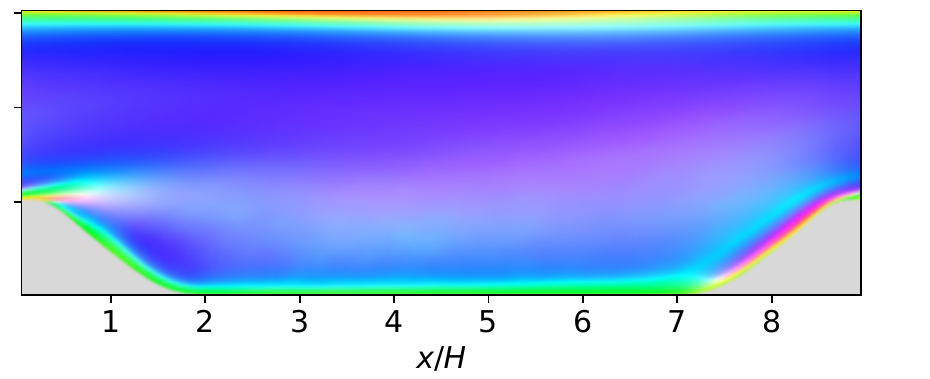}
        \caption{DNS}
    \end{subfigure}
    \vspace{-0.1cm}
    \caption[Visualization of stress types of the $k-\omega$ model, the TBNN with different feature sets, and DNS data for periodic hills.]{Visualization of stress types of LEVM $k-\omega$ \textbf{(a)}, 
    PI-TBNN \textbf{(b)}, and DNS \textbf{(c)}}
    \label{fig:prediction_b_ph_rgb_map}
\end{figure}

The PI-TBNN was tested on flow cases it had not seen during training.
They differ either in geometry or Reynolds number from the training data.
Figure \ref{fig:prediction_b_sd_rgb_map} compares the anisotropy tensors for square duct (SD) using the barycentric colormap.
Only the TBNN with the extended feature set can accurately reproduce the state of turbulence for this flow case.
A similar picture arises on the PH geometry in Figure \ref{fig:prediction_b_ph_rgb_map}, where the $k-\omega$ model cannot capture 1C turbulence and axisymmetric expansion at the top and the bulk of the flow domain, respectively.
The PI-TBNN, however, does exhibit such characteristics.
For all test cases considered, the PI-TBNN achieves about 70\% reduction of the RMSE compared to the baseline $k-\omega$ model and 50\% reduction of the RMSE compared to \cite{Ling2016} (see Table \ref{tab:rmse_b}).

\begin{table}
    \centering
    \footnotesize
    \caption[Root-mean-square error of $\bs{b}$ from RANS and PI-TBNN predictions for square duct flow.]{RMSE of $\bs{b}$ from RANS, PI-TBNN, and TBNN predictions for the three test cases.}
    \begin{tabular}{p{3.65cm} p{1.75cm} p{1.9cm} p{2.35cm} p{2.1cm}}
        \toprule[0.9pt]
        Flow case & LEVM $k-\omega$ & PI-TBNN, FS1 & PI-TBNN, FS[1-3] & {TBNN Ling \cite{Ling2016}} \\
        \midrule[0.4pt]
        Square duct & $0.2175$ & $0.0992$ & $0.0663$ & $0.14$ \\
        Periodic hills & $0.1016$ & $0.0628$ & $0.0419$ & \\
        Curved backward-facing step & $0.1173$ & $0.0619$ & $0.0414$ & \\
        \bottomrule[.9pt]
    \end{tabular}
    \vspace{0.2cm}
    \label{tab:rmse_b}
\end{table}

\subsection{Anisotropy Tensor Propagation}

\begin{figure}
    \vspace{0.25cm}
    \centering
    \begin{subfigure}[b]{0.49\textwidth}
        \includegraphics[width=\textwidth]{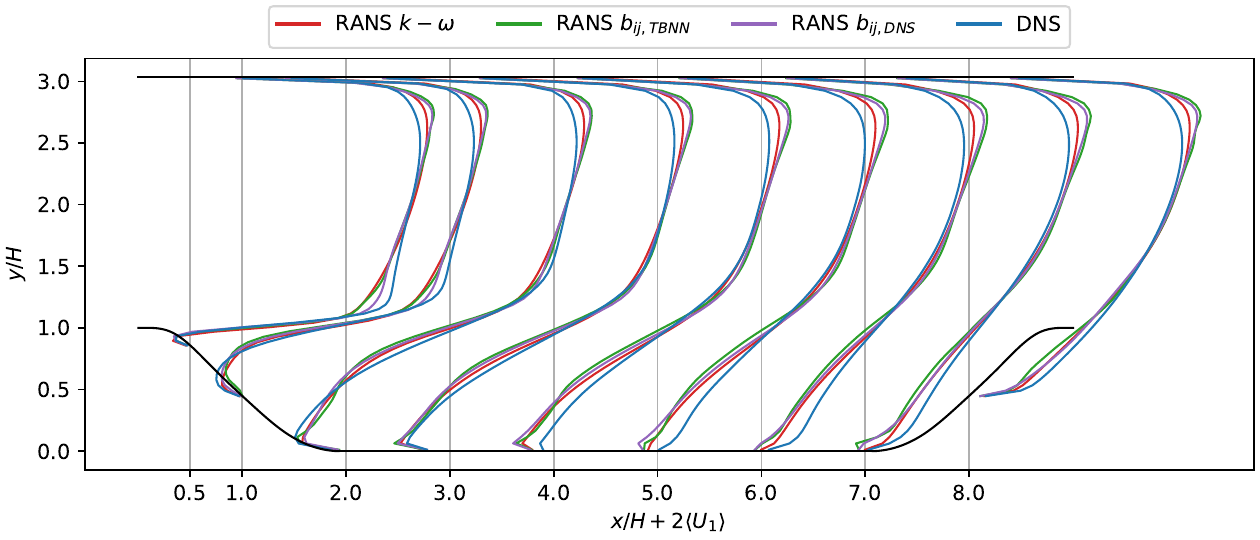}
        \caption{Periodic hills}
    \end{subfigure}
    \begin{subfigure}[b]{0.49\textwidth}
        \includegraphics[width=\textwidth]{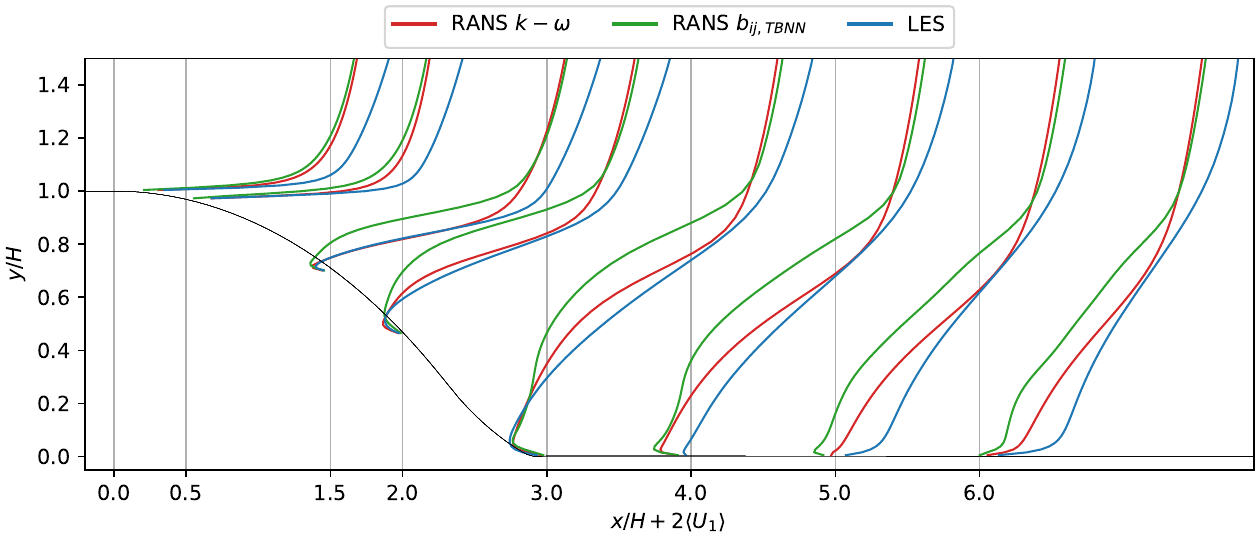}
        \caption{Curved backward-facing step}
    \end{subfigure}
    \caption[Streamwise mean velocity profiles at specific $x$-locations of the periodic hills flow case.]{Streamwise mean velocity profiles at specific $x$-locations of PH \textbf{(a)}, and CBFS \textbf{(b)}.}
    \label{fig:propagation_ph_cfbs_section}
\end{figure}

While some applications involving wall shear stress computations may directly benefit from an improved prediction of $\bs{b}$, the quantities of interest are usually the mean velocity and pressure fields.
Hence, with the PI-TBNN model, the Reynolds equations were solved for the mean velocity and pressure fields.
Table \ref{tab:rmse_u} shows that the PI-TBNN outperforms the $k - \epsilon$ model for the PH and CBFS geometries by a small margin and yields better in-plane prediction for the square duct flow case.
The most accurate model, however, remains the $k-\omega$ model for all three test geometries.
Surprisingly, on the CBFS geometry, the PI-TBNN shows the largest discrepancies in the region of the flow separation, as can be seen in Figure \ref{fig:propagation_ph_cfbs_section}.
The $k-\omega$ model even beats the mean field resulting from using the anisotropy tensor from the DNS on the periodic hills test case, indicating that an improved anisotropy tensor does not necessarily lead to improved mean velocity and pressure fields (behavior also reported in \cite{Taghizadeh2020TurbulenceConsiderations,Duraisamy2021PerspectivesTurbulence}).
Both \cite{Ling2016} and \cite{Wu2018} reported improvements in the mean-field prediction over a LEVM but used the $k - \epsilon$ as the baseline LEVM, which shows a larger discrepancy from the ground truth than the $k - \omega$ model, i.e. the best performing model of its class.

\begin{table}
\centering
\footnotesize
\caption{RMSE of $\bs{U}$ for RANS with anisotropy tensor from $k-\omega$, $k-\epsilon$, PI-TBNN, and DNS. Reference velocity fields come from DNS.}
\begin{tabular}{p{5.2cm} p{1.7cm} p{1.7cm} p{1.7cm} p{1.3cm}}
\toprule[0.9pt]
 Flow case & $k-\omega$ & $k-\epsilon$ & $\bs{b}_{\mathrm{PI-TBNN}}$ & $\bs{b}_{\mathrm{DNS}}$\\
\midrule[0.4pt]
Square duct RMSE$(\bs{U})$ & $0.0496$ & $0.0667$ & $0.0716$ & $0.0322$\\
Square duct RMSE$([U_2, U_3])$ & $0.0066$ & $0.0066$ & $0.0045$ & $0.0025$\\
Periodic hills RMSE$(\bs{U})$ & $0.0375$ & $0.0545$ & $0.0541$ & $0.0465$\\
Curved backward-facing step RMSE$(\bs{U})$ & $0.0609$ & $0.0883$ & $0.0868$\\
\bottomrule[.9pt]
\end{tabular}
\label{tab:rmse_u}
\end{table}

\section{Conclusion}

We introduced the PI-TBNN, which extended the TBNN framework with an extensive feature set and an inductive bias in the form of a physics-informed addition to the loss function. The addition of features was motivated both analytically---showing that the number of distinct invariants for 2D flow scenarios is three, not five---and empirically through improved predictions. It has been shown that the new approach yields more accurate predictions of the anisotropy tensor than the original TBNN of Ling et al. \cite{Ling2016}. The improvements were illustrated with the barycentric colormap and quantified by comparing RMSEs. It is, however, limited in its predictive capabilities of the mean velocity and pressure fields. While it still outperformed the widely popular $k-\epsilon$ model on geometries with flow separation, it consistently fails to compete with the $k-\omega$ model.
The rather large discrepancy of the RANS using the DNS anisotropy tensor indicates that it is more beneficial to train models that not only aim at improving predictions of the Reynolds stresses but instead target the mean field quantities directly, e.g., as in \cite{Agrawal2023AUncertainty, Hayek2018, Parish2016}.

\section{Broader Impact}

Turbulence is a key physical characteristic of a broad range of fluid flows. Understanding this phenomenon is crucial for complex designs, environmental modeling, and many more engineering applications. Computational power has increased massively in the past decades, enabling scale-resolving simulations like direct numerical simulations of a number of canonical turbulent flows. However, fast approximations like the RANS continue to remain essential for industrial applications, whose accuracy hinges heavily on turbulence closure models.

The presented research serves as an extension to the state-of-the-art data-driven turbulence closure model proposed by \cite{Ling2016}. By combining a deep neural network with an inductive bias informed by turbulence realizability constraints, plus an extensive feature set, the PI-TBNN showed considerable improvements. These improvements showcased through barycentric colormap visualizations and quantified reductions in RMSE signify a step forward in our ability to capture complex turbulence states, particularly in scenarios involving surface curvature and flow separation.

However, the study also sheds light on the nuanced relationship between improved anisotropy tensor predictions and the ultimate goal of predicting mean velocity and pressure fields. While the PI-TBNN excels in enhancing anisotropy tensor predictions, it does not consistently outperform the established $k-\omega$ model in mean-field predictions, underscoring the need for further exploration in this area.

We do not see any direct ethical concerns associated with this research. The impact on society is
primarily through the over-arching context of research using machine learning to improve our general understanding of turbulence in fluids.



{
\small
\sloppy
\bibliography{references}
}



\section{Supplementary Material}


\subsection{RGB colormap}
\label{subsec:colormap}

As demonstrated in Figure \ref{fig:rgb_map_scaled}, each point in the barycentric triangle corresponds to a unique color. The mapping from the barycentric coordinates to the RGB values follows

\begin{align}
\label{eq:rgb_map_scaled}
    \begin{bmatrix}
        \mathrm{R} \\ \mathrm{G} \\ \mathrm{B}
    \end{bmatrix}
    = \frac{1}{\max C_{ic}} \left(C_{1c} \begin{bmatrix}
        1 \\ 0 \\ 0
    \end{bmatrix}
    + C_{2c} \begin{bmatrix}
        0 \\ 1 \\ 0
    \end{bmatrix}
    + C_{3c} \begin{bmatrix}
        0 \\ 0 \\ 1
    \end{bmatrix} \right) \quad \mathrm{for} \; i \in \{1,2,3\}.
\end{align}

\subsection{Scalar invariants}
\label{sec:invar_proof}
\noindent Two of the five scalar invariants ($\lambda_3, \lambda_4$) are zero for flows with one direction of homogeneity. The two invariants read

\begin{align}
    \lambda_3 = \mathrm{tr}(\bs{\hat S}^3), \hspace{3cm}
    \lambda_4 = \mathrm{tr}(\bs{\hat \Omega}^2 \bs{\hat S}).
\end{align}

\noindent Assuming the flow is homogeneous in $z$-direction, the partial derivatives of the mean velocity with respect to $z$ vanish, and the mean rate of strain and rotation read

\begin{align}
    \hat S_{ij} = \frac{1}{2} \frac{k}{\epsilon}
    \begin{bmatrix}
2 \deriv{\mean{U_x}}{x}   & \deriv{\mean{U_x}}{y} + \deriv{\mean{U_y}}{x}\\
\deriv{\mean{U_y}}{x} + \deriv{\mean{U_x}}{y} & 2 \deriv{\mean{U_y}}{y}\\
\end{bmatrix},\hspace{0.25cm}
\hat \Omega_{ij} = \frac{1}{2} \frac{k}{\epsilon}
    \begin{bmatrix}
0   & \deriv{\mean{U_x}}{y} - \deriv{\mean{U_y}}{x} \\
\deriv{\mean{U_y}}{x} - \deriv{\mean{U_x}}{y} & 0\\
\end{bmatrix}.
\end{align}

\noindent The incompressibility constraint of the Reynolds equations reduces to

\begin{align}
    \deriv{\mean{U_i}}{x_i} = \mathrm{tr}\left( \deriv{\mean{U_i}}{x_j} \right) = \deriv{\mean{U_x}}{x} + \deriv{\mean{U_y}}{y} = 0.
\end{align}

\noindent When using the simplified expressions of $\bs{\hat S}$ and $\bs{\hat \Omega}$ in combination with the incompressibility constraint, invariant $\lambda_3$ is given by

\begin{align}
    \mathrm{tr}(\hat S_{ij}^3) &=\\
    \footnotesize{\frac{k^3}{8\epsilon^3}}
    \mathrm{tr} &\left( {\footnotesize \begin{bmatrix}
        \hat{S}_{11} (\hat{S}_{11}^2 + \hat{S}_{12}^2) + \hat{S}_{12}(\hat{S}_{11}\hat{S}_{12} + \hat{S}_{12}\hat{S}_{22})  
        & \hat{S}_{12}(\hat{S}_{11}^2 + \hat{S}_{12}^2) + \hat{S}_{22} (\hat{S}_{11}\hat{S}_{12} + \hat{S}_{12}\hat{S}_{22}) 
        \\
        \hat{S}_{11} (\hat{S}_{11}\hat{S}_{12} + \hat{S}_{12}\hat{S}_{22}) + \hat{S}_{12}(\hat{S}_{12}^2 + \hat{S}_{22}^2) 
        & \hat{S}_{22} (\hat{S}_{12}^2 + \hat{S}_{22}^2) + \hat{S}_{12}(\hat{S}_{11}\hat{S}_{12} + \hat{S}_{12}\hat{S}_{22}) 
        \\
    \end{bmatrix}} \right) \nonumber\\
    &= \frac{k^3}{8\epsilon^3} \left( \hat{S}_{11} (\hat{S}_{11}^2 + \hat{S}_{12}^2) + 2\hat{S}_{12}^2\underbrace{(\hat{S}_{11} + \hat{S}_{22})}_{=0} 
    + \hat{S}_{22} (\hat{S}_{22}^2 + \hat{S}_{12}^2) \right), \\
    &= \frac{k^3}{8\epsilon^3} \left( \hat{S}_{11} (\hat{S}_{11}^2 + \hat{S}_{12}^2) + \hat{S}_{22} (\hat{S}_{11}^2 + \hat{S}_{12}^2) \right),\\ 
    &= \frac{k^3}{8\epsilon^3} \left( \underbrace{(\hat{S}_{11} + \hat{S}_{22})}_{=0} (\hat{S}_{11}^2 + \hat{S}_{12}^2) \right) = 0.
\end{align}

\noindent The derivation of invariant $\lambda_4$ is more straightforward and thus written in terms of the mean velocity gradient. It is given by

\begin{align}
    \mathrm{tr}(\hat \Omega_{ij}^2 \hat S_{jk}) &= \mathrm{tr}\left( \frac{k^3}{8 \epsilon^3} \left(\deriv{\mean{U_x}}{y} - \deriv{\mean{U_y}}{x} \right)^2
        \begin{bmatrix}
            2 \deriv{\mean{U_x}}{x}   & \deriv{\mean{U_x}}{y} + \deriv{\mean{U_y}}{x} & 0 \\
            \deriv{\mean{U_y}}{x} + \deriv{\mean{U_x}}{y} & 2 \deriv{\mean{U_y}}{y} & 0 \\
            0 & 0 & 0 \\
        \end{bmatrix}
    \right) \\
    &= \frac{k^3}{4 \epsilon^3} \left(\deriv{\mean{U_x}}{y} - \deriv{\mean{U_y}}{x} \right)
    \underbrace{\left( \deriv{\mean{U_x}}{x} + \deriv{\mean{U_y}}{y} \right)}_{=0} = 0.
\end{align}

\subsection{Data set}
\label{subsec:data_set}

The high-fidelity direct numerical simulation (DNS)/large eddy simulation (LES) data are used for the training. Out of the two available DNS for the CDC, the one at ${\mathrm{Re}=7900}$ was used for estimating the regularization parameters.
The higher Reynolds number simulation at ${\mathrm{Re}=12600}$ was used for training and validation, along with the DNS at ${\mathrm{Re}=2800}$ and LES at ${\mathrm{Re}=10595}$ for PH and the DNS for SD at $\mathrm{Re} = \{2000, 2400, 2900, 3200\}$.
The data set was split into training and validation sets at a ratio of $70/30$. Therefore, the total number of data points available for training was $26600$.



The testing set consists of three flow geometries (SD, PH, and CBFS).
Two of these geometries, SD and PH, are also part of the training and validation set, however, at different Reynolds numbers. The periodic hills case at ${\mathrm{Re} = 5600}$ was selected to investigate the interpolation properties --- the ML model has previously seen this flow geometry and is expected to yield good predictions. The square duct is a canonical flow case that clearly illustrates the deficiencies of the LEVMs and is suitable to present difficulties of propagating the Reynolds stresses to the flow field. Finally, the curved backward-facing step tests the ML model's extrapolation capabilities.

\end{document}